\documentstyle[psfrag,preprint,eqsecnum,amsfonts,floats,aps,tighten,epsfig]{revtex}

\def\Pl{\ell_P}
\def\lm{L}
\def\be{\begin{equation}}
\def\ee{\end{equation}}
\def\g{{\gamma}}
\def\a{\alpha}
\def\b{\beta}

\def\d{\delta}
\def\l{\lambda}

\def\A{{\cal A}}

\def\H{{\cal H}}

\def\G{{\cal G}}
\def\Ab{\overline{\A}}

\def\S{\Sigma}
\def\W{{\cal W}}

\def\ba{\begin{eqnarray}}
\def\ea{\end{eqnarray}}

\def\C{{\cal C}}
\def\lp{\ell_P}

\newcommand{\brckt}[1]{\langle #1 \rangle}

\newcommand{\braket}[2]{\langle #1 \mid #2 \rangle}

\newcommand{\pic}[5]{\raisebox{#3pt}
{\hspace{#4pt} \epsfig{file=#1.eps,height=#2pt,silent=} 
\hspace{#5pt}}}
\newcommand{\kd}[1]{\mathchoice{
\pic{#1}{24}{-8}{-1}{2}}{
\pic{#1}{11}{-3}{1}{1}}{
\pic{#1}{9}{-2}{-3}{1}}{
\pic{#1}{7}{-1}{-1}{0}}}

\begin{document}

\draft

\title{A Gaussian Weave for Kinematical Loop Quantum Gravity}
\author{A. Corichi${}^{1}$\thanks{Electronic address: 
corichi@nuclecu.unam.mx}
and
J. M. Reyes${}^{1,2}$\thanks{Electronic address:
mreyes@fis.cinvestav.mx}
} 

\address{${}^{1}$Instituto de Ciencias Nucleares, 
Universidad Nacional Aut\'onoma de M\'exico\\ 
A. Postal 70-543, M\'exico D.F. 04510, M\'exico.} 

\address{${}^{2}$Departamento de F\'{\i}sica, 
Centro de Investigaci\'on y de Estudios Avanzados 
del IPN\\
A. Postal 14-740, M\'exico D.F. 07000, M\'exico.}
\date{\today}
\maketitle

\begin{abstract}
Remarkable efforts in the study of the semi-classical regime of 
kinematical loop quantum gravity are currently underway. In this note,
we construct a ``quasi-coherent'' weave state
using Gaussian factors. In a similar fashion to
some other proposals, this state is 
peaked in both the connection and the spin network basis. 
However, the state constructed here has the novel feature that, 
in the spin network basis, the main contribution for 
this state is given by the fundamental representation, independently
of the value of the parameter that regulates the Gaussian width.
\end{abstract}
\pacs{}

\section{Introduction}
\label{sec1}

One of the most important physical 
results of loop quantum gravity \cite{review} is the prediction
that the spectra of geometric operators, 
corresponding to their classical analogues such as areas of surfaces 
and volumes of regions, are purely discrete \cite{rs,AL}. 
Some other intriguing features of this `Non-perturbative
Quantum Geometry' have emerged already at the kinematical level,
for example, the non-commutativity of geometric
operators \cite{ACZ}, and the statistical derivation of Black Hole
entropy \cite{ABCK}.
Despite the relative success of 
the canonical/loop quantum gravity approach,
two major problems need to be successfully
tackled: dynamics (that is, a complete 
and anomaly-free implementation
of the quantum Hamiltonian constraint) 
and the recovery of general relativity as a low energy/macroscopic
regime of the quantum theory. How could the two problems be related is an 
open question, but it is likely that a complete understanding 
of the  quasi-classical
regime of the theory at the kinematical level 
(in particular states which can 
approximate functions of connections) can give some insight into the
dynamics of the theory \cite{VZ}.

In this approach, General Relativity (GR) has to arise from the 
semi-classical limit of the quantum theory through properly 
defined  `coherent states' and a suitable coarse-graining.
Coherent states have been constructed only recently in loop 
quantum gravity \cite{thiemann1}. However, we shall argue that
some important issues have to be 
addressed first in order to obtain a complete understanding
of the problem.
At the kinematical level, the first attempts in this direction were
given by
the so called `weave states' in the old loop representation \cite{ARS}.
As originally constructed, weave states were intended to be 
semi-classical states (initially,
eigenstates of geometrical operators) that approximate smooth 
geometries on a background spatial (and compact) 3-manifold $\S$
for large scales. That is, given a fiducial metric  on $\S$, the 
spatial average of expectation values of area and volume operators in
this state should be given by their classical values as measured
by the fiducial 3-metric on $\S$ above some macroscopic scale $\lm$. 
There is increasing
hope that once one has a good candidate for a weave describing flat
space, one would be able to develop quantum field theory (QFT) 
on these states. There are some evidences pointing out in the direction 
that QFT's on weaves would be free of infinities and should not 
require any renormalization procedure \cite{thiemann2}.

Another important issue that 
deserves further attention is whether semi-classical states 
can be promoted to solutions to
the Hamiltonian constraint of the theory. It is well known that 
eigenstates of 3-geometries are not suitable states
describing 4-dimensional Minkowski space-time, as happens,  for instance,
in quantum electrodynamics (QED) where an eigenstate of the electric field
operator
with null eigenvalue has a very different physical meaning than that of
QED vacuum \cite{IR}. However, kinematical results could be important
by themselves
in situations where classical boundary conditions such as asymptotical 
flat space times \cite{AG} and isolated black holes boundary conditions
are imposed on the system \cite{VZ,ABCK}.

Recent attempts in constructing weaves were given in
\cite{GR} with the use of `spin networks', and have been recently 
used as a background `thermal' state to 
construct new Hilbert spaces for non-compact spaces \cite{AG}. 
Nevertheless, weaves are defined based on the behavior of 
geometrical operators
that depend only in momentum variables 
and therefore, only provide  information about half of the 
classical phase space while the behavior of some connection variable 
operators might not be under control in these states. In fact, 
the original weaves were 
expected to be `concentrated' in geometry, 
so they are highly delocalized in the
connection variable and the fluctuations of the
configuration operators might be very large. For instance,
it has been recently shown that
the natural strategy for approximating connections
by taking the elementary configuration
variable, namely the $SU(2)$ holonomy, as the 
multiplicative operator, fails \cite{VZ}. 
Any attempt to reproduce their classical
analogue is doomed since there are no quantum
states that support such supposition and such that they
satisfy additional minimum uncertainty relations \cite{VZ}. 
Instead, a more ambitious proposal has been given by the
authors of \cite{VZ} and consists on a tentative  
definition of  `quasi-classicity',
a latticization of the underlying manifold and the use of 
`magnetic flux' type operators. This 
was the first step in attempting to construct
within the loop approach, realistic semi-classical states
at the kinematical level (in the simplified case of 
two spatial dimensions). More recently, a
 series of papers by Thiemann and
Winkler have appeared in which semi-classical states
are constructed from a (complex) coherent-state transform 
on phase space (see \cite{thiemann1} and references therein).

In this work we follow the same strategy as in \cite{AG}, namely we
construct  an (invertible) weave 
state based on a cylindrical function, that might 
serve as background for the construction of new Hilbert spaces
when the space is non-compact.
There is however, an important difference between our
weave state and the one originally proposed in \cite{AG}:
Our state depends in a `Gaussian way', i.e. on a suitable exponential
arrangement of square of group elements. 
The state constructed here is peaked around the trivial `flat'
connection that yields the identity element for the holonomy around
each loop defining the state. With this prescription, one might 
hope to overcome the objection previously raised regarding large
fluctuations for (every) holonomy operator. In this sense, if one
is able to reproduce the flat (metric) behavior through the
geometric operators, and the connection is also `peaked' around the
flat connection, one might hope to have a semi-classical state
approximating Minkowski space-time.

It is worth mentioning that
the exponential dependence on group elements 
is a  common feature
of coherent states in compact Lie groups \cite{Perelomov}.
We will show that this one-parameter family of
states has  better behavior than the original 
`quasi-coherent' weave, and that in this case,
the states are peaked and centered around the 
fundamental representation, when written in the spin network basis.
This result is particularly intriguing,
since a similar behavior, i.e., a dominance of the
fundamental representation, is also observed in the computation
of the black hole entropy \cite{ABCK}.

This paper is organized as follows: In Section \ref{sec2} we give a
very brief review of the most important facts of loop quantum gravity that
are relevant for this work. Readers familiar with the loop formalism
may want to skip to Section \ref{sec3} where the notion of a weave state
is recalled. In the Section \ref{3.2} we proceed 
to the construction of the
`Gaussian' weave which approximates flat space. 
We also take the opportunity
to make some general observations and, in particular,  we
discuss some subtleties that arise
when an eigenbasis of the area operator is introduced. 
We close with a discussion in Section \ref{dis}.
 
\section{Preliminaries}
\label{sec2}

Canonical/Loop quantum gravity is a canonical 
approach for quantizing general relativity in a non-perturbative way.
In this approach, general relativity is a purely constrained 
system, a feature for all diffeomorphism invariant theories.
A rigorous kinematical framework to handle theories whose 
classical configuration space is given by connections 
{\em moduli} local gauge transformations,
has been constructed in the past decade \cite{los5}. 
General relativity can be 
written in this form using the so-called Ashtekar-Barbero
variables \cite{A}. In this
section  we summarize the main results of this approach.

The classical phase space consists of canonical pairs of
a $SU(2)$ valued connection $A_a^i$ on an orientable spatial 3-manifold 
$\S$ and its conjugate momentum variable, a densitized triad $E_i^a$ 
which takes values in the dual of the Lie algebra \cite{A}. The 
dynamics of the theory, for compact $\S$,
is pure {\em gauge} and is encoded in the
Gauss constraint which generates $SU(2)$ gauge transformations,
the spatial 3-dimensional diffeomorphism constraint  
generating  spatial 3d diffeomorphisms on $\S$, 
and the Hamiltonian constraint which generates  the 
coordinate time evolution.

The classical configuration space $\A$, is given by the space 
of all smooth connections on a 
principal $SU(2)$ bundle over $\S$. Since $\A$ is infinite dimensional
the quantum configuration space $\Ab$  is a certain
completion of the classical one which includes all `distribution-like' 
connection,  in a similar role played in free field theory in
Minkowskian space by the tempered distributions.
 Hence $\Ab$ is taken to be the space of 
generalized connections which have a well defined action on 
oriented paths over $\S$ and the group of generalized 
$SU(2)$ gauge transformations \cite{los5}.
 
Since $\Ab$ is compact it admits a regular diffeomorphism-invariant 
measure $d\mu_o$ 
and the Hilbert space can be taken to be the space 
$\H=L^2(\Ab,d\mu_o)$ 
of square-integrable functions on $\Ab$. This space is actually 
non-separable, but we can associate quantum states of 
the gauge theory to finite graphs $\g$ \footnote{For simplicity we 
will take the paths to be analytic embedded.}. If we consider 
all possible graphs, we obtain a very large set of states 
which are {\it dense} on $\H$. 

These states defined on finite graphs
are called {\it cylindrical functions} and depend on finite sets of 
holonomies along the edges of the graph. Therefore, given any complex
valued function $\psi:(SU(2))^n\rightarrow{\bf C}$ we can associate 
a cylindrical function $\Psi_\g(A)\in\A_\g$ as follows:
\be
\Psi_{\g,\psi}(A)=\psi(h{}_{e_{1}}(A),..,h{}_{e_{n}}(A)). \label{cf}
\ee
$h{}_{e_{i}}$ denote the $SU(2)$ parallel propagator matrices of the
connection $A \in \Ab$ along the curve $e_i$. Exploiting the compactness
of $SU(2)$ the space of cylindrical functions $\C$ 
is equipped with the inner product:
\be
\braket{\Psi_{\g_1,\psi_1}}{\Psi_{\g_2,\psi_2}}= \int_{SU(2)^n}
\overline{\psi_1(h{}_{e_{1}},..,h{}_{e_{n}})}
\psi_2(h{}_{e_{1}},..,h{}_{e_{n}})
d \mu_H(h{}_{e_{1}}) \ldots d \mu_H(h{}_{e_{n}})
\ee
where $d \mu_H(h{}_{e_{1}}) \ldots d \mu_H(h{}_{e_{n}})$ is the n-copy
Haar measure naturally induced from $SU(2)$. If the graphs $\g_1$ and $\g_2$
are different, the respective cylindrical functions can be viewed as being 
defined in the same graph, which is the union of both: $\g_1 \cup \g_2$.

The classical algebra of elementary observables $\cal{S}$ consist of the 
set of canonical variables $A_a^i$ and $e_{abi}:=\eta_{abc}E^c_i$ 
smeared against suitable 
fields which in this case  are one dimensional for the 
connection, and two-dimensional for the triad field \cite{ACZ}. 
The space of configuration variables
is given by the space Cyl of functions of the type (\ref{cf}), when
the graph $\g$ is varied over all arbitrary finite graph. 
The momentum variables are obtained by smearing 
the pseudo 2-forms $e_{abi}$ against a test field $f^i$ which takes 
values in  the Lie-algebra of $SU(2)$. 
\be
^2E[S,f] = \int_S e_{abi}f^idS^{ab}.
\ee
Here, the surface $S$ is  restricted to be of the type 
$S=\overline{S}-\partial \overline{S}$, where
$\overline{S}$ is any compact, analytic 2-dimensional sub-manifold of $\S$ 
possibly with boundary. See \cite{ACZ} for details and \cite{thiemann7}
for an alternative choice of variables.
The standard representation of the quantum algebra on ${\cal H}$ is to have the
cylindrical (configuration) functions act as multiplicative operators, and
``electric flux'' variables as certain derivative operators \cite{ACZ}.

When the spatial slice $\S$ is non-compact, a quantum state that describes,
say, an asymptotically-flat space requires an infinite volume (and
a contribution to this volume coming from everywhere within $\S$). This
in turn, implies that the state must have contributions from
graphs with an infinite number of edges. This state would not, 
in particular,  belong to our Hilbert space ${\cal H}$.
In order to have a formalism that would allow for these situations,
Arnsdorf and Gupta have proposed an algebraic construction, a la
GNS, to overcome this difficulty. The basic idea is to construct
new Hilbert spaces where a particular, semi-classical, non-normalizable
state would serve as a vacuum of the theory, and the ordinary
normalizable states would be seen as `finite excitations' of the
geometry. For details of the construction see \cite{AG}.
For our purposes, it is enough to know that the formalism requires
an invertible state, that is, a state that does not vanish for any
point on $\Ab$, and that has support on an infinite number of edges.
In the next section we shall construct such an state.

\section{Approximating flat space}
\label{sec3}

This section has two parts. In the first one we recall the general
definition of a weave state. In the second part, we construct a
Gaussian weave and show that it satisfies the requirements
for a decent weave state and that it has some interesting features that
make it an acceptable candidate for a quantum theory on non-compact 
spaces.

\subsection{General Framework}

In any quantum field theory, the semi-classical regime is obtained or
represented by two main ingredients: ``coherent states''  and a suitable 
coarse graining. However, in the case of the gravitational field 
additional problems arise since the notion of ``vacuum state'' is not
provided a priori, unless one forces it by splitting the metric into 
a preferred vacuum metric (usually taken flat) and a perturbative
term whose fluctuations define the quantum theory. In the other hand, 
in canonical/loop quantum gravity, as in any other non-perturbative
approach to quantum gravity, the natural `vacuum state' corresponds to 
the `no metric' space $g_{ab}=0$. In this case flat space becomes a
highly excited state of the 3-geometry, containing an infinite
number of fundamental excitations.  
Weave states are kinematic states constructed to
provide semi-classical metric information over the spatial 3-manifold 
$\S$, that is, they 
approximate flat space above some macroscopic scale $L$. 
However, since
the spatial metric $g_{ab}$ is constructed  completely from 
momentum variables (densitized triads), weaves are normally
thought to be highly 
concentrated in geometry, and it is possible that certain
configuration variables,  when promoted
to operators, would be wildly delocalized in these states.
This would render the weave states
as unsuitable candidates for quasi-classical states.
However,
the weave state that we shall construct here is peaked, in the
connection representation, around the flat connection for
each loop of the defining graph. Thus, one expects that
certain configuration observables be localized on these
states.
It is important to recall that all relevant results in the
semi-classical regime of the theory have emerged at the 
kinematical\footnote{wherein the diffeomorphism and  Hamiltonian
constraint are ignored.} level since a complete and anomaly-free
implementation of the constraints have not been found
successfully.

The discrete nature of quantum geometry
\cite{rs,AL,ACZ,RDP} provides a suitable criterion for approximating 
flat space. There are geometric operators
which are self-adjoint in the Hilbert space $\H$ and that
measure areas of surfaces and volumes of regions. Therefore, a
classical spatial metric can be approximated by requiring that the
expectation values of areas and volumes of macroscopic surfaces
and regions correspond to their classical analogues.

Let suppose that we want to approximate the Euclidean 3-space 
$R^3$ with the weave state $\Phi_w$. The strategy is as follows: Let consider
a macroscopic region in $R^3$ (of size larger than $\lm$) with bulk $R$ 
and surface $S$. We perform successive measurements of the volume
of the region ${\hat V}_{R}$ and the area of the surface ${\hat A}_{S}$
on the state $\Phi_w$ at different positions in $\S$.
The state $\Phi_w$ is called  a `weave' state
if it satisfies the following requirements:

\begin{enumerate}
	\item The spatial average of the expectation values of the 
	area $S$ and volume $R$ agree with their classical values.
		\begin{mathletters}
		\label{expvalues}
		\be
		\overline{\brckt{ {\hat A}_{S}}}_w= 
		\int_{S}\sqrt{\mid q_{ab}\mid } + O(\epsilon^2/\lm^2),
		\ee
		\be
		\overline{\brckt{ {\hat V}_{R}}}_w= 
		\int_{R}\sqrt{\mid g_{ab}\mid } + O(\epsilon^3/\lm^3),
		\ee
		\end{mathletters}
	where $\lm$ is the characteristic macroscopic size of the
	bulk and $\epsilon$ is a microscopic scale, which is normally
taken to be of the order of the Planck length $\lp$, $q_{ab}$ is the induced metric on $S$, the ``over-line'' indicates spatial average and the expectation values are 
	taken in the state $\Phi_w$.
	
	\item The state $\Phi_w$ must have small quantum fluctuations in the
	measurements with regard to the macroscopic scale $\lm$.
		\begin{mathletters}
		\label{deviations}
		\ba
		\d_V \ll \lm^3 \\
		\d_A \ll \lm^2 
		\ea
		\end{mathletters}
	where $\d_V$ and $\d_A$ are the standard deviations of the 
	repeated measurements of the volume and area of the 
	corresponding region.
\end{enumerate}

Actually, the conditions (\ref{deviations}) can be satisfied by an infinitely
number of quantum states and do not single out a preferred state.
In general, one could identify two distinct steps in the construction
of a candidate of a weave state. First, one specifies a graph on $\S$ where
the state is supposed to have support (or a family of  graphs in a 
statistical sum). For instance, one could provide a lattice that `covers'
the whole space $\S$ \cite{VZ}, or a `unit' graph $\gamma$ that is going to
be spread over $\S$ to cover it. The second step involves the specification of
a cylindrical function to be living on the basic units of the construction,
be it on plaquetes of the lattice in the first case above, or basic
cylindric functions on the graph $\gamma$.
Some  explicit examples of both types
have been constructed in the literature, the
first ones being simultaneous eigenstates of the volume and area
operators  (\cite{GR},\cite{ARS}) and a recent `quasi-coherent' state
\cite{AG} with some nice properties. An example of states based on a lattice
is given in \cite{VZ}.

In this note, we construct a particular weave state similar in spirit to the
one given in \cite{AG}. Actually, one is only proposing a cylinder function
(i.e., the second step in the construction outlined above), which could
in principle be implemented on different graphs. We choose a graph construction
first proposed in \cite{GR} (and also considered in \cite{AG}) for
simplicity, and to be able to compare the properties of our Gaussian
ansatz with the other proposals. It should be clear from the discussion that
one could take our Gaussian ansatz and implement it in other graph/lattice
constructions. Our ansatz has the property that it shares some of the 
nice properties of earlier proposals, that is: it is an invertible state, is 
peaked in the connection and the spin-network basis. However, it 
has the particular
feature that the peak in the spin-network basis is centered around 
the fundamental  representation. 
We shall refer to this state as the `Gaussian' weave because it 
is constructed as a Gaussian function of cylindrical factors.

\subsection{The `Gaussian' Weave}
\label{3.2}

\begin{figure}
\begin{center}
\epsfig{file=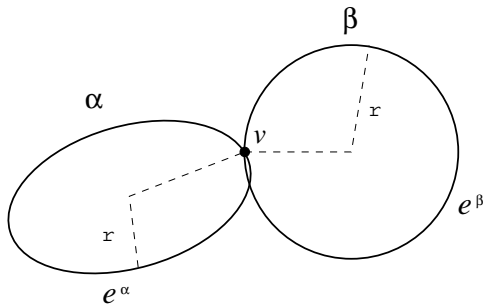,width=2.5in,silent=}
\end{center}
\caption{Graph $\g_i$ as the union of two intersecting circles
$\a_i$ and $\b_i$. Note that the intersection is given by a 
tetravalent vertex.}
\label{lens}
\end{figure}

The aim of a weave is to reproduce the area and volume of {\em any} region
in the slice $\S$. Therefore, the graph in which the state is based must fill
all $\S$ and if the space $\S$ is non-compact, one is left with two options: The 
graph either consists of edges of infinite length or it is an infinite 
set of finite graphs. In particular,
our weave state is based on a disjoint union of an infinite 
collection of finite graphs: $\G_\rho^r=\cup_{i=1}^{\infty}\g_i$ where
$\g_i$ represent the graph of a system of two non-coplanar circles $\a_i$
and $\b_i$ with the same radius $r$ respect to a fiducial metric, 
which intersect
in a single vertex, see Fig.\ \ref{lens}. These systems of circles 
are randomly
distributed in $\S$ with average density $\rho$ measured with the 
same fiducial metric.

Let us define the cylindrical function based in the graph $\g_i$
\be
g_i(A)=N \exp{-(\lambda {\rm Tr}[(h_e{}_i^\a (A)h_e{}_i^\b(A)-e)_{(1)}])^2},
\label{fg}
\ee
where $N$ is a normalization factor, $\lambda$ is a positive
parameter, $h_e{}_i^{\a,\b} (A)$ are the holonomies along the circles
$\a$ and $\b$ respectively and the subindex $1$ represents the 
fundamental representation in $SU(2)$ (Color 1 in the terminology of
Penrose). It is worth noting the Gaussian 
dependence on group elements. The Gaussian weave can be written as an 
infinite product of functions (\ref{fg}).
\be
\W=\lim_{n \rightarrow \infty} \prod_{i=1}^n g_i(A).  \label{we}
\ee
This state does not belong to the Hilbert space $\H$ but 
is defined through the cylindrical function $g_i$ based on the 
graph $\g_i$ and can be considered as a candidate for a background (thermal)
state for constructing new Hilbert spaces as in \cite{AG}.

The state (\ref{we}) has common features with that given in \cite{AG}
and can be considered as a `quasi-coherent' state in the same footing. 
By definition, the cylindrical functions given by the finite product
$\W_n=\prod_{i=1}^n g_i(A)$
take on their maximum values when the holonomies
along the circles in the graph $\g_i$ are the identity in $SU(2)$ and if $A_e$
denote the connection which give the trivial holonomy, as 
$n \rightarrow \infty$
the function $\W_n$ become increasingly peaked around $A_e$. The width of the 
peak is regulated by the Gaussian parameter $\l$.

We are interested in exploring the main features of the state $\W$ and in
showing that it satisfies the weave conditions (\ref{expvalues}), and 
(\ref{deviations}). For this purpose we begin expanding
the cylindrical function $g_i$ into the De-Pietri-Rovelli
spin-network basis. The only elements of the basis that have a non-vanishing
contribution are given by, 
\be
\Phi_j={\rm Tr}[(h_e{}_i^\a (A)h_e{}_i^\b(A))_{(j)}],
\label{basis}
\ee
based on the graph $\g_i$, where the subindex $j$ denote the `color' 
representation
of $SU(2)$. This choice arises because $\Phi_j$ can be written as a 
linear expansion of spin-network
states {\em some} of which are eigenstates of the area operator for
an arbitrary analytic surface $S$ that intersect the graph $\g_i$.
The fact that these states are not eigenstates of the area operator
for {\it any} surface is
particularly interesting and deserves further explanation. 

On the full Hilbert space $\H$, and for gauge-invariant cylindrical functions 
$N_\g$\footnote{We will restrict to gauge-invariant states on $\Ab$. 
This will be 
useful because it permits the transition to the loop representation 
where from the
beginning one restrict oneself to gauge invariant states 
(spin-network states).},
the spectrum of the area operator $\hat{A}$ is unbounded
from above and the `area gap', which is the smallest non-zero 
eigenvalue, is given in 
the special situation when the graph intersect the surface $S$ with 
an -up or down-
edge and a tangential edge at a bivalent vertex $v$ \cite{AL}. 
If a single edge 
intersect the surface (without crossing $S$), or if the edges that 
meets at the vertex $v$ are {\em all} `up' or `down', the 
Gauss law constraint implies that the corresponding
eigenvalue be $0$. Therefore, to obtain non-trivial eigenvalues 
of the area operator is necessary that the edges of the graph 
`cross' the surface $S$. This means that for 
the system of circles $\g_i$, the state (\ref{basis}) is an 
eigenstate of the area operator with non-trivial eigenvalues 
if at least one of the circles crosses $S$, that
is, if the same circle intersect the surface in two points (vertices). 
If one or both of the circles only `touch' the surface in one point, 
the arcs that meet that point will be at the same side of 
$S$ and the corresponding eigenvalue would be $0$ as was stated above. 
The maximum number of intersections between $\g_i$ and the
surface is $4$ and is obtained when both circles cross $S$. 

\begin{figure}
\begin{center}
\epsfig{file=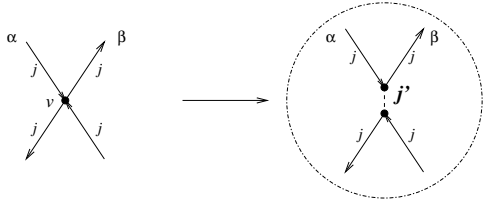,width=2.5in,silent=}
\end{center}
\caption{Decomposition of the tetravalent vertex into two
trivalent vertices joined by an 'internal' edge of color $j'$ inside 
the virtual ribbon-net.}
\label{4v}
\end{figure}

On the other hand, given the spin-network $s=(\g_i, \vec{j}, \vec{l})$ 
which assigns
the labelling $\vec{j}=(j,j)$ to the circles $\a_i$ and $\b_i$ of the 
graph $\g_i$
with the $j$ representation of $SU(2)$, and a labelling 
$\vec{l}=(l_1)$ of the single tetravalent vertex of $\g_i$ with
 an invariant tensor
$c_1$, that is,  an intertwining tensor from the representations of 
the incoming
edges to the representations of the outgoing edges. The spin-network 
states can be
constructed by contracting the matrices associated to the $j$ 
representation of the
circles on $\g_i$ with the intertwining tensor $c_1$ at the 
tetravalent vertex. The basis element
(\ref{basis}) when written in the spin-network (state) expansion, 
correspond to the
choice of a particular basis of the space of the invariant tensor of 
the tetravalent vertex,
and is given by the Fig.\ \ref{4v} in the 
De Pietri-Rovelli notation \cite{RDP}.

Graphically, the choice
of the basis in the space of invariant tensors is equivalent 
to the `decomposition' of 
the tetravalent vertex into a trivalent graph  with two vertices 
joined by a `virtual' edge $j'$. Since the Clebsh-Gordan condition 
must hold in each of the trivalent vertices, 
the only allowed value for the virtual vertex is $j'=0$. Thus, 
we would be lead to the conclusion that the basis
element (\ref{basis}) 
is an eigenstate of the area operator. This is
true in general, but it remains the subtle case when the 
tetravalent vertex of the
graph $\g_i$ {\em lies} on the surface $S$. There are two possible
(non-generic) configurations in  this special case:

\begin{enumerate}
\item The first configuration is obtained when the surface 
$S$ divides the graph
$\g_i$ into two halfs. Each of them has a part of the circle $\a_i$ 
{\em and} a part
of the circle $\b_i$. In this situation, (\ref{basis}) has no
contribution to the area coming from the vertex (since $j'=0$) and
is then an eigenstate of the area
operator with eigenvalues proportional to
$8 \pi \Pl^2 \sqrt{\frac{j}{2}(\frac{j}{2}+1)}$
for each intersection of the circles with $S$. 
	
\item The other `non-generic' configuration is realized when the 
surface $S$ divides
the graph $\g_i$ in such way that one of the circles, say 
$\a_i$, is entirely placed
on one side of the surface and the circle $\b_i$ 
on the other. In this case the
basis (\ref{basis}) is {\em not} an eigenstate of 
the area operator and it is 
indispensable the use of the recoupling theorem 
\cite{RDP,KAU} in order to perform the suitable change of basis in 
which the area operator can take a diagonal form. Even when the
state is written in a diagonal form, there will be in general contributions
from eigenstates with different eigenvalues for the area.
\end{enumerate}

At first sight this seems to be a problem when expanding the Gaussian function
(\ref{fg}) into the 
basis elements (\ref{basis}) but a closer look shows that 
the number of
`pathological' configurations represents a zero measure set in the collection 
of all possible
configurations. However, even though the configurations with the 
tetravalent vertex on
$S$ do not contribute in a significant way to the expansion of the 
function (\ref{fg}), they are 
at the same time, conceptually relevant. The fact that not every 
spin-network state is
an eigenstate of the area operator, in the case when the vertex has 
valence four or more, was noted earlier  
\cite{ACZ2}, but seems not to be widely recognized in the
literature. This particular feature leads to the conclusion
that the area operators of two surfaces that intersect along a 
line fail to commute for 
states which have 4 or higher valent vertex in the intersection. 
Thus, the quantum Riemannian geometry that arises 
from loop gravity is 
intrinsically non-commutativity \cite{ACZ}. Therefore, some questions 
arise concerning the 
semi-classical regime of the theory: Under what conditions do
 we expect to obtain a
commutative geometry in the semi-classical approach?  If the
 notion of the weave
is a good candidate to describe classical space, What are the 
conditions on the weave
states in order to reconcile our classical notion of a 
commutative geometry? These issues are under current study.

Once we have shown that it is plausible to expand the cylindrical 
function (\ref{fg})
into the basis (\ref{basis}) we proceed to determine the coefficients of the 
linear expansion 
$g_i(A)=\sum_{j}^{} c_j\Phi_j$ to be able to evaluate the 
expectation values of the area $\langle\hat{A}\rangle$, the volume
$\langle\hat{V}\rangle$ and their corresponding deviations. Let us
start by defining the cylindrical function $g_i$ by its series expansion
\be
g_i=N e^{-\l^2(\Phi_1-2)^2}=N e^{-4 \l^2} \sum_{n=0}^{\infty}
\frac{\l^n H_n(2 \l)}{n!} \Phi_1^n.
\ee
$H_n(x)$ denote the Hermite polynomials of order $n$. This expression can be
reduce noting that the following tensorial-algebra relation holds in theory of
representations of $SU(2)$.
\be
\Phi_1^n=\sum_{j}^{}a_j^n\Phi_j
\qquad{\rm where} \qquad
a^n_j  =  \frac{(j+1)n!}{(\frac{n-j}{2})!(\frac{n+j}{2}+1)!} \qquad
   \mathrm{for} \ \frac{n-j}{2} \in Z^+, \label{su2}
\ee
and $a^n_j = 0$ otherwise. Defining 
$\C(\lambda)=Ne^{-4\lambda^2}$ the coefficients $c_j$ of the expansion
can be written as
\be
c_j=\C(\lambda)\sum_{n=0}^{\infty}\frac{\lambda^nH_n(2\lambda)}{n!}a_j^n, 
\nonumber
\ee
Thus,  using the relations (\ref{su2}) the cylindrical function (\ref{fg}) 
expanded on the basis $\Phi_j$ is given by
\be
g_i(A)=\C(\lambda)\sum_{j}^{}(j+1)\sum_{n=0}^{\infty}
\frac{\lambda^nH_n(2\lambda)}{(\frac{n-j}{2})!(\frac{n+j}{2}+1)!}
 \Phi_j, \label{gphi}
\ee
We would want to evaluate the sum in the index $n$ of the equation (\ref{gphi})
but it can not be done exactly. Therefore, one is forced to use the 
definition of Hermite
polynomials in its series expansion an evaluate the expression numerically. The
coefficients $c_j$ are given by 
\be
c_j=\C(\lambda)(j+1)\sum_{k=0}^{\infty}\frac{\lambda^{2k+j}}{k!(k+j+1)!}
\sum_{m=0}^{[\frac{2k+j}{2}]}\frac{(-1)^m(2k+j)!}{m!(2k+j-2m)!}
(4\lambda)^{2k+j-2m}, \label{coeff}
\ee
where $[\nu]$ denotes the larger integer $\leq\nu$.

\begin{figure}
\begin{center}
\psfrag{colors j}{$\text{\small{colors}}j$}
\epsfig{file=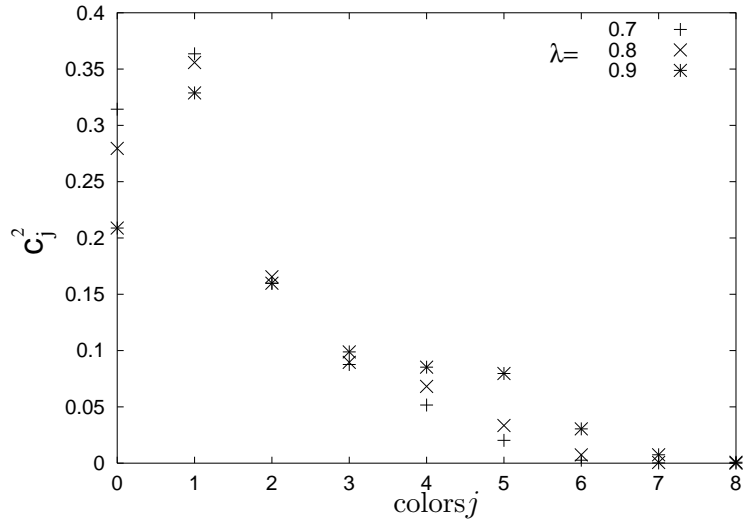,width=4in,silent=}
\end{center}
\caption{Contribution coefficients of the state in the spin 
network basis for some values of the parameter $\lambda$. Note that 
the peak is centered at the fundamental representation.}
\label{cc}
\end{figure}

A numerical evaluation of equation (\ref{coeff}) was performed and shows, with
the normalization $\sum_{i=0}^{\infty}c_j^2=1$, that the series 
converges in the interval $0<\lambda<1$ and for values above 
$\lambda\geq0.5$ a peak starts to grow at the fundamental 
representation. Fig.\ \ref{cc} shows some coefficients for some
values of $\lambda$. It is important to stress out that for larger values
of the 'color' $j$ the expansion continues, but in this case
a numerical evaluation turns out to be inappropriate due to float
point errors.

The most important and intriguing difference between the `quasi-coherent' weave
\cite{AG} and the `Gaussian'  weave is that, in the last case the 
peak of the coefficients
(\ref{coeff}) in the basis of spin-network states is centered 
around the fundamental
representation $j=1$. The dominance of the fundamental representation in other
physical situation such as the BH entropy seems to point to a deep
role played by the fundamental representation in loop quantum 
gravity. The origin of this behavior deserves further attention.

One can get a closed formula for the expectation value for the 
area in the state
$g_i$ using the machinery of recoupling theory \cite{RDP,KAU} in the loop 
representation
\be
\brckt{\hat A}=\sqrt{\frac{j}{2}(\frac{j}{2}+1)}
\sum_{k}^{}\sum_{j}^{}c_{k}^{*}c_j
\braket{\kd{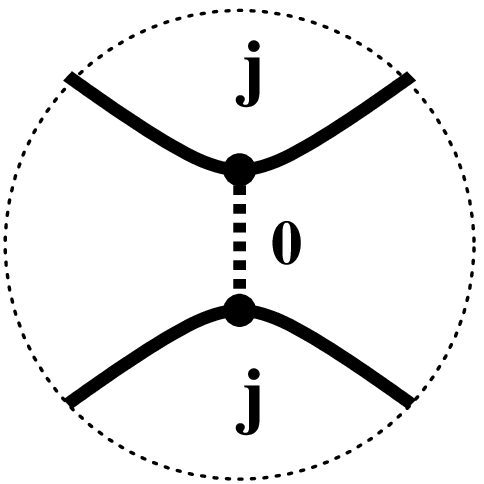}}{\kd{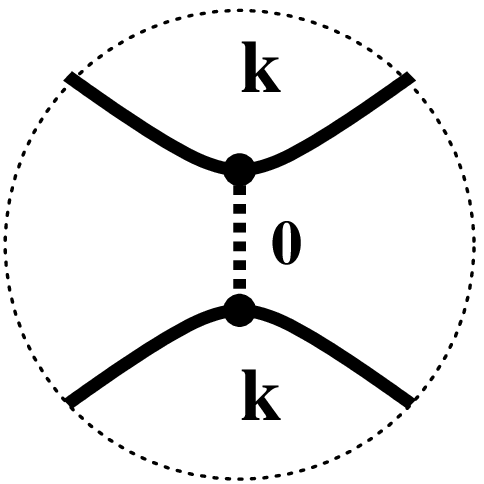}}.
\ee
Where the bracket can be evaluated chromatically to show that the
basis states are normalized.

Thus, the expectation value of the area operator is
given by,

\be
\brckt{\hat A}=(8\pi\Pl^2)\sum_{j}^{}|c_j|^2
\sqrt{\frac{j}{2}(\frac{j}{2}+1)}.
\ee

In the case of the volume operator, a closed formula for the 
expectation value is not available. 
However, we know that the largest eigenvalue in the state $\Psi_j$ 
increase as $j^{3/2}$ and we can make an estimation. 
For the particular value of 
the Gaussian width $\lambda=0.75$ we have the following 
expectation values for the
geometric operators and their respective dispersions:
\begin{mathletters}
\label{evaluations}
\ba
\langle \hat{A}\rangle=1.149(8\pi\Pl^2) \qquad ;\qquad
\Delta_A:=\sqrt{\langle \hat{A}^2\rangle-\langle \hat{A}\rangle^2}=
0.432(8\pi\Pl^2),  \\
\langle \hat{V}\rangle=1.078(8\pi\Pl^2)^{3/2} \qquad; \qquad
\Delta_V:=\sqrt{\langle \hat{V}^2\rangle-\langle \hat{V}\rangle^2}=
2.435(8\pi\Pl^2)^{3/2},
\ea
\end{mathletters}
As expected, the mean values and dispersion of the basic 
geometric observable are of the order of $\Pl^2$ in the case of the 
area and of $\Pl^3$ in the case of the volume. Our state $g_i$ is 
peaked in area and volume as in a similar fashion as the
proposal given in \cite{AG}, but with better accuracy. 
It is easy to see that the state (\ref{we}) satisfies the 
necessary conditions to be considered
a weave state, because the expectation 
values (\ref{evaluations})
are of the order of suitable powers of the Planck length. 
For a formal demonstration we  refer to \cite{AG}.
\section{Discussion and outlook}
\label{dis}

In this work we have proposed a basic cylindrical function
with a Gaussian dependence on the group element.
This function can be used, in particular, to construct a
weave state a la Grot-Rovelli. The particular proposal has the
desired features of a weave state plus a peakedness property
around the fundamental representation.

There exists a striking similarity between the main 
contribution of the state
in the fundamental representation and the dominating 
states, in the statistical counting,
that contributes to the entropy of an 
`isolated horizon'. This dominant
contribution correspond to punctures
all of which have labels in the fundamental representation 
(spin $1/2$) \cite{ABCK}. 

The Gaussian state proposed here can play an important role both 
as a background  to construct new Hilbert spaces
and as a suitable state in the recent proposal for
lattice based semi-classical states
in loop quantum gravity  \cite{VZ}.

We finish by pointing out several issues that deserve further investigation:
\begin{enumerate}

\item One would like to understand the physical foundations for the 
fact that there is a dominance of
the fundamental representation in at least two different situations.

\item It is necessary to investigate the  role played by these (or other) 
`quasi-classical' states in  the
recovery of a commutative classical geometry. 

\item In the construction of new Hilbert spaces a la GNS, different choices
of `vacuum states' might lead to unitarily inequivalent Hilbert spaces.
It is important to understand if the Gaussian ansatz presented
here yields an equivalent or inequivalent quantum theory to the one 
constructed in \cite{AG}.  There are also coherent states constructed using
`infinite tensor product Hilbert spaces' \cite{thiemann1}. 
The relation of this method
to the algebraic GNS  construction remains unclear and deserves further 
attention.

\item Finally, we would like to understand the relation of the Gaussian ansatz
to the coherent states constructed in \cite{thiemann1}, specially since 
both are constructed using coherent state-like states on the group.

\end{enumerate}

Some of these issues are under investigation and shall be reported elsewhere.

\acknowledgements

We would like to thank S. Gupta and J.A. Zapata for discussions, and a
referee for helpful comments. This 
work was partially funded by DGAPA No. IN121298 and CONACyT No.
J32754-E grants. JMR was supported in part by CONACyT scholarship No. 85976.

\end {document}